\documentclass[12pt]{iopart}
\usepackage{citesort,epsfig}
\begin{document}

\title{Formation of a metastable phase due to the presence
of impurities}
\author{Richard P. Sear\\
Department of Physics, University of Surrey,
Guildford, Surrey GU2 7XH, United Kingdom,
{\tt r.sear@surrey.ac.uk}}

\begin{abstract}
Phase transitions into a new phase that is itself
metastable are common; instead of the
equilibrium phase nucleating a metastable phase does so.
When this occurs the system is sometimes said to be obeying
Ostwald's rule.
We show how this can happen when there
are impurities present that reduce the barrier to heterogeneous
nucleation of the metastable phase. We do so by studying
a Potts lattice model using Monte Carlo simulation.
Thus, which phase forms
depends not only on the properties of the different phases
but also on the impurities present. 
Understanding why systems obey Ostwald's rule may therefore require
a study of the impurities present.
\end{abstract}

\maketitle

\section{Introduction}

On heating glassy silica it crystallises into a crystalline
form called cristoballite, but the equilibrium crystalline
form is tridymite \cite{kingery}. The silica has two crystalline
forms and it transforms into the metastable form, not the equilibrium form.
Phase transformations into metastable phases are
quite common, and systems that do this
are sometimes said to obey the Ostwald or Ostwald step
rule \cite{wolde99}.
Typically the appearance
of the less stable phase is ascribed to it being in some
way more similar to the original phase it nucleated in
than is the equilibrium phase and so
having a lower interfacial tension with the original phase.
Within classical nucleation theory the nucleation rate
is proportional to $\exp(-\gamma^3/h^2)$, where $\gamma$ is the interfacial
tension between the nucleating phase and the phase it is nucleating
in and $h$ is proportional to the difference in chemical potential
between the nucleating phase and the phase
it is nucleating in \cite{kingery,debenedetti}.
Thus, although $h$ will be larger for the nucleation of the equilibrium
phase, if the interfacial tension $\gamma$ for this phase
is also larger its rate of nucleation may be slower
than that of a metastable phase.

This argument is based on the nucleation of the new phases
being homogeneous, i.e., occurring in the bulk. However,
the nucleation of most new phases
is not homogeneous but heterogeneous, it takes place in contact
with impurities, or with the surface of the
container \cite{kingery,nucon,cantor,perepezko}.
This implies that which phase nucleates first may be influenced
by differences in the interactions of the nuclei of the
different phases with the impurities present.
Here, we use computer simulation and theory to
demonstrate that indeed impurities can determine which phase
appears.
We find that if we start with a system in which
the equilibrium
phase nucleates, we can change only the nature
of the impurity and obtain a system in which the metastable
phase nucleates.

We study nucleation in the Potts lattice
model \cite{wu82,janke97}, via Monte Carlo computer simulation. This model
is one of the simplest models that has the three phases we require.
Thus we will try to obtain an understanding of the generic
features of nucleation under conditions
when there is more than one phase that
is more stable than the existing phase and when there are impurities
present. We hope our conclusions
will apply widely to systems where there are competing
phases that the system can transform into.
The nucleus of a new phase is typically
only a few molecules (in the case of the Potts model: spins) across and so
even impurities only a few molecules or spins across
are large enough to greatly
reduce the free-energy barrier to nucleation; see for example
\cite{cacciuto04}. Here we will study impurities only a few
spins across, although we could have studied much larger impurities.
We defer a systematic study of the effect of varying the impurity
size to later work. See e.g.,~\cite{toxvaerd02} and references
therein for recent simulation work on heterogeneous nucleation.

In the next section we define a simple model that has the required
three phases.
The third section contains the results of Monte Carlo simulations.
These simulations are exact and demonstrate that the nature
of the impurity can indeed determine which phase nucleates. Having
established this, we do not go on and systematically vary
parameters such as the size, shape etc.~of the nucleus. Instead,
in the fourth section we write down a simple phenomenological theory
for the competitive nucleation of two phases, one of which
is assisted by an impurity, i.e., the nucleation is heterogeneous,
and the other of which is not and so nucleates via homogeneous
nucleation. This allows us to calculate how much an impurity needs
to reduce the barrier to nucleation in order to determine which
phase nucleates, as a function of the density of impurities, the
interfacial tensions between the phases etc. The final section
is a conclusion, where we discuss the relevance to experiment.

\section{Potts model}

Consider the three-state Potts model \cite{wu82}
on a simple cubic
lattice in three dimensions.
On each lattice site $i$ there is a spin $s_i$ that can take
one of three values: 1, 2 or 3, and that
interacts with its six nearest neighbours. The interaction
energy of a pair of neighbouring spins $i$ and $j$
is $-J\delta_{s_is_j}$, i.e., the only interaction is between
spins that have the same spin value.
$J$ is positive, so on cooling
the model undergoes a symmetry-breaking transition
from a state in which a third of the spins have each spin value
to one of three ordered phases, in each of which one of the spin values
predominates \cite{wu82}. These are the spin-one, spin-two
and spin-three phases.
In the absence of any external
fields all three phases have the same free energy.
The transition occurs at $J/kT=0.55$ \cite{janke97}.
$k$ and $T$ are Boltzmann's constant and the temperature,
respectively. Here we work solely at the low temperature
$J/kT=0.8$.

We will consider not the disorder-order
transition but transitions between the three ordered phases.
To do so we need
to consider external fields
that break the symmetry between these phases.
The three external
fields $h_k$, $k=1,2,3$, couple to the spins via terms
$-h_k\delta_{s_ik}$, i.e., a positive $h_k$ favours the
phase with spins predominantly taking the value $k$.
If, for example, the $h_k$ have values
$h_3>h_2>h_1$, then the spin-three phase is the equilibrium phase
and the spin-two phase is more stable than the spin-one phase.
We will always start in the spin-one phase and always set $h_1=0$.
Then by increasing $h_2$ and $h_3$ from zero we will make
both the spin-two and spin-three phases more stable than
the spin-one phase.

\begin{figure}[t]
\caption{
\lineskip 2pt
\lineskiplimit 2pt
Computer simulation snapshot of a small
simulation box 20 spins across.
The system is
in the spin-one phase at coexistence, $h_1=h_2=h_3=0$,
at a temperature $J/kT=0.8$.
Sites with spins taking the values two and three are filled with
green and very pale cubes, respectively. The fixed spins that
form the impurity are dark red. The impurity favours the spin-two
phase, it is a $p=2$ impurity, and it is a square monolayer
with sides $8$ spins long.
Sites with free spins taking the value one are left empty.
\label{snap}
}
\vspace*{0.3in}
\begin{center}
\epsfig{file=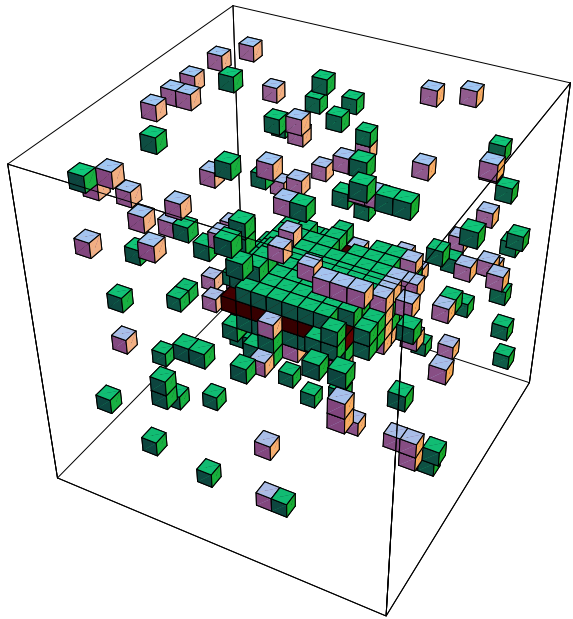,width=7.5cm}
\end{center}
\end{figure}

We also require an impurity that favours
either the spin-two or the spin-three phase. We use an impurity that is
a square monolayer of $8$ by $8$
spins that are fixed and that interact with adjacent
spins with an energy $-2J\delta_{s_ip}$, where
when $p=2$ ($p=3$) we have an impurity with a strong affinity
for the spin-two (spin-three) phase. See figure \ref{snap}
for a snapshot showing the impurity.
We refer to the spins
that can flip between the three values as free spins to
distinguish them from the fixed spins that form the impurity.
At the temperature we work at the interaction between the
fixed spins of the impurity and the free spins is strong enough
that if the impurity is expanded into an infinite plane wetting
occurs. See for example \cite{schick,bonn01} for an introduction to wetting.
For example, if $p=2$, at equilibrium at
coexistence, $h_1=h_2=h_3=0$, impurities in either
spin-one or spin-three phases are wet by the spin-two phase.
In between the, infinite, impurity and the bulk phase there will
be a macroscopic layer of the spin-two phase
and then an interface between the spin-two phase
and either the spin-one or spin-three phase.
We verified this via computer simulation.

Putting all the interactions together the energy
of a configuration of the spins is
\begin{equation}
H=-J\sum'_{ij}\delta_{s_is_j}-\sum_{k}h_k\sum_i\delta_{s_ik}
-2J\sum''_{i}\delta_{s_ip}.
\label{hamil}
\end{equation}
The first, dashed,
sum is over all nearest-neighbour pairs of
free spins. The two sums in the middle terms are over the
three applied fields and over all spins.
The double-dashed
sum in the last term is over all free spins adjacent to the
impurity.

\section{Simulation results}

\begin{figure}[t]
\caption{
\lineskip 2pt
\lineskiplimit 2pt
Plot of the fractions of spins taking the values two (top set
of points) and three (bottom set). A point is plotted
every cycle, i.e., one attempted flip per spin.
The simulation is stopped after 1881 cycles as then the
spin-two phase has nucleated and grown so that 45\% of the spins
have the value two.
The number of spin-three fluctuates: on average a little less
than 2\% of the spins are spin-threes, this corresponds to around
500 spin-threes. This relatively small number fluctuates. The number
of spin-two fluctuates much more.
\label{s23}
}
\vspace*{0.3in}
\begin{center}
\epsfig{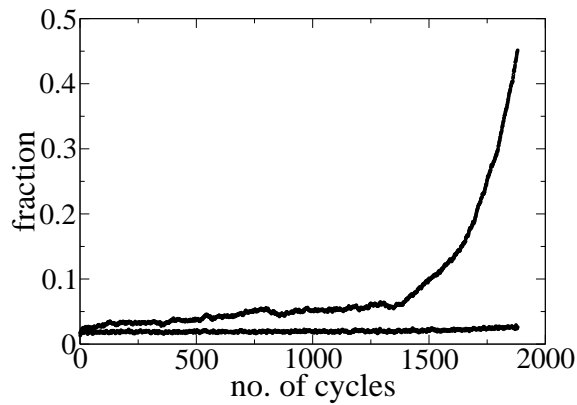}
\end{center}
\end{figure}

We simulate using the standard Metropolis Monte Carlo method for spins.
Each move starts by selecting one of the free spins at random. This
spin is then flipped to either of the two other spin states
with equal probability. If this flip lowers the energy it
is always accepted, if it increases the energy it is accepted
with a probability that is the exponential of minus negative
of the energy change over $kT$.
See \cite{chandler} for an introduction to the Monte Carlo
method. Our simulations were done on a lattice 
of 30 by 30 by 30 spins with periodic boundary conditions.
This is a somewhat larger lattice than shown in the snapshot
of figure \ref{snap}. The appearance of a new phase was determined
by monitoring the fraction of spin-twos and spin-threes.
Once the fraction of spin-twos exceeded 45\% the simulation was stopped
and the spin-two phase was taken to have nucleated.
Similarly,
if the fraction of spin-threes exceeded 45\% the spin-three
phase was taken to have nucleated.
See figure \ref{s23} for a plot of the fractions of spins that were
spin-twos and spin-threes as a function of simulation time, for
one simulation run.
The fraction 45\% is arbitrary,
varying it even by large amounts does not change the result in almost
all cases. Note that then there are over ten thousand spins in the
new phase so the nucleus is clearly post-critical, it
is extremely unlikely that its growth will stop.
In most cases we repeated the simulation five times with the
same values of $h_2$ and $h_3$, but near the borderlines between
different nucleation behaviours we performed ten simulation runs.
If neither the spin-two nor the spin-three phase nucleated within
100,000 cycles, the simulation was abandoned. There the
nucleation rate is too low to be measured via direct simulation.

We start all our simulations
by setting $h_1=h_2=h_3=0$ and then equilibrating the system
in the spin-one phase. For our first simulations we then
instantaneously increased both $h_2$ and $h_3$ to $0.2$.
The spin-two and spin-three phases are then equally stable,
and more stable than the spin-one phase.
We found that if the simulation box contains
a $p=2$ impurity, which has an affinity for the spin-two phase,
then the spin-two phase nucleates, whereas if the box contains
a $p=3$ impurity,
then the spin-three phase appears \cite{variant}.
If $h_3$ is increased
to $0.3$, keeping $h_2=0.2$, then with a $p=2$ impurity the
spin-two phase still appears not the spin-three phase, even
though the spin-three phase is now more stable. As we expect
if at $h_2=0.2$, $h_3=0.3$, the $p=2$ impurity is replaced
by a $p=3$ impurity it is the spin-three phase that appears.
So, it is the impurity that is controlling which phase appears.
This is the key result of this work.
In Table \ref{table1} along
the diagonal, i.e., $h_2=h_3$, and also
even for $h_2=h_3-0.1$, it is always the spin-two phase that appears.
However, returning to a system with a $p=2$ impurity, if while
$h_2$ is kept at $0.2$, $h_3$ is further increased to $0.5$, it
is the spin-three phase that nucleates.
As the stability of the equilibrium phase with respect
to that of the metastable phase is increased, by increasing
$h_3-h_2$, beyond a certain limit our impurity no longer
controls the phase that forms.
For $h_2=0.2$ and $h_3=0.5$,
$p=2$ and $p=3$ impurities both result in the spin-three phase
forming. Results for a number of different values
of $h_2$ and $h_3$, are given
in table \ref{table1}. These results are for a $p=2$ impurity.
Of course, due to the symmetry
of the model, results for a $p=3$ impurity can be obtained by
simply swapping the labels 2 and 3 in table \ref{table1}.

\begin{table}
\caption{Results of computer simulations of a system of $30^3$
Potts spins at $J/kT=0.8$ with a $p=2$ square impurity $8$ spins along
each side.
Starting with a system equilibrated in the spin-one
phase at $h_1=h_2=h_3=0$ the system has $h_2$ and $h_3$
instantaneously increased to the values shown.
Formation of the spin-two phase is indicated by a `2' and formation
of the spin-three phase by a `3'. A `0' indicates that
neither phase nucleated within 100,000 cycles.
The `2/3' indicates that of 10 simulation runs 3 resulted
in the spin-two phase and 7 in the spin-three phase. Here the barriers
to formation of these two phases are comparable.
Finally, $3^*$ indicates that both the spin-three and the
spin-two phase nucleated but that
it was the fraction of spin-threes that grew to exceed
45\%.
}
\begin{center}
\begin{tabular}{|c|c|c|c|c|}
\hline
$h_2=$ & 0.1 & 0.2 & 0.3 & 0.4 \\
\hline
$h_3=$ & & & &  \\
0.2 & 0 & 2 & 2 & 2 \\
0.3 & 0 & 2 & 2 & 2 \\
0.4 & 3 & 2/3 & 2 & 2  \\
0.5 & 3 & 3 & $3^*$ & 2  \\
\hline
\end{tabular}
\end{center}
\label{table1}
\end{table}

Stranski and Totomanow \cite{stranski33} argued that when a
system is in a phase
which has a higher free energy than more than one
other phase the phase that nucleates is the one with the lowest
nucleation barrier.
We agree, but as nucleation
is typically heterogeneous, which phase has the lowest
barrier will depend on what impurities are present, as
well as on properties of the phase itself.
As an example, consider increasing the size of the impurity.
If the impurity is expanded into an infinite plane, then it will be
wet by the spin-two phase. Thus, if the impurity is large enough
then the spin-two phase will nucleate effectively at $h_2=0$
as already at this value of $h_2$ a macroscopic wetting layer
of the spin-two phase
will be present on the surface of a macroscopic impurity.
Thus, if such a large impurity is present the spin-three
phase will only get a chance to nucleate
in the spin-one phase if the spin-two phase is actually
higher in free energy than the spin-one phase, $h_2<0$.
Note that if $h_3$ is sufficiently large the spin-three phase
may of course nucleate from the spin-two phase.

\section{Phenomenological theory for the competitive nucleation
of two phases}

A sufficiently large impurity with a surface wet by a
new phase can reduce the nucleation
barrier to zero, even at coexistence, where the barrier
to homogeneous nucleation is 
divergent \cite{debenedetti}. Alternatively, if a dilute impurity
does not prefer the nucleating phase it will not participate
in nucleation.
Denoting the barriers to homogeneous and heterogeneous
nucleation by $\Delta F^*_{HOMO}$ and $\Delta F^*_{HET}$,
respectively, the magnitude
of the difference $\delta$ is then defined by
$\Delta F^*_{HET}=\Delta F^*_{HOMO}-\delta$; it is
the free energy difference between the critical nuclei.
For a sufficiently large impurity with a surface that is wet
when the surface is infinite, $\delta$ can be made arbitrarily
large. It can also be effectively zero. Note that $\delta$ will
in general be a complex function of $h_2$ and $h_3$ because
impurities will change the shape and size of the critical
nucleus.

We would like to explore the nucleation behaviour
for the parameter
space composed of the driving forces
for nucleation of the spin-two and spin-three
phases, $h_2$ and $h_3$, and the effect of the impurity,
$\delta$. For simplicity, we will
only consider $p=2$ impurities that favour the spin-two phase,
and we will neglect homogeneous nucleation of the spin-two phase.
Thus we will consider only the competition between homogeneous
nucleation of the spin-three phase and heterogeneous nucleation
of the spin-two phase. It is easy to relax this constraint but
it introduces additional variables without changing the qualitative
nature of the behaviour.
Exploring the parameter space via computer
simulation would be very laborious but fortunately classical
nucleation theory \cite{debenedetti} should be accurate enough
for this purpose. Classical nucleation theory has been shown
to be very reasonable for homogeneous nucleation in the Ising model well below
this model's critical temperature, where the transition is strongly
first-order as it is here, see for example \cite{wonczak00}
and references therein.

We will start with the classical nucleation theory \cite{debenedetti}
for homogeneous
nucleation of the spin-three phase in the spin-one phase.
Although this is a little inaccurate we consider nuclei to be
always perfectly cubic, i.e., to consist of $\lambda$ by
$\lambda$ by $\lambda$ spins. The free energy change on
forming a nucleus is just the sum of a bulk
term from creating a volume $\lambda^3$ of the spin-three phase and
a surface term from creating $6\lambda^2$ of spin-three--spin-one
interface. Then the free energy of a nucleus
of the spin-three phase is \cite{debenedetti}
\begin{equation}
\Delta F_3=-\lambda^3h_3+6\lambda^2J,
\label{homo}
\end{equation}
where we used the low-temperature approximation for the interfacial
tension between the spin-one and spin-three phases
$\gamma\simeq J$. The rate is determined by the free energy
of the nucleus, equation (\ref{homo}), at the top of the barrier,
which is
$\Delta F^*_3=32J^3/h_3^2$.
The rate of homogeneous nucleation of the spin-three
phase, per lattice site, $r_3$, is then \cite{debenedetti}
\begin{eqnarray}
r_3=\nu\exp\left[-32J^3/(h_3^2kT)\right],\nonumber\\
\label{r3}
\end{eqnarray}
where $\nu$ is an attempt frequency; it is of the same
order as the frequency
of spin flips at a site. The rate of homogeneous nucleation
of the spin-two phase is just that given by equation (\ref{r3})
with the field $h_2$ replacing $h_3$. The interfacial
tension between the spin-one and spin-two phase will be very similar
to that between the spin-one and the spin-three phases at these
low temperatures. Then, according to the definition of $\delta$
the barrier to heterogeneous nucleation of the spin-two
phase is $\Delta F^*_2=32J^3/h_2^2-\delta$.
The rate of heterogeneous nucleation of the spin-two phase, per lattice site,
$r_2$, is therefore
\begin{eqnarray}
r_2
&=&\nu\rho_i\exp\left[-32J^3/(h_2^2kT)+\delta/kT\right],
\label{r2}
\end{eqnarray}
where $\rho_i$ is the number of impurities divided by the number
of sites. We expect the density of impurities to be very low so
we fix $\rho_i=10^{-6}$. Also, for an impurity of specific
material, size etc., then $\delta$ will be a function of $h_2$
and $h_3$. We ignore this dependence here and treat $\delta$
as simply a shift in the nucleation barrier.
Having determined how large a shift is needed we can then work
back to estimate the properties the impurity must have
in order to generate it.

\begin{figure}
\caption{
\lineskip 2pt
\lineskiplimit 2pt
Plot showing the conditions where the spin-two phase nucleates first,
above the curve, and, where the spin-three
phase nucleates first, below the curve.
Each of the two curves gives the value of $\delta$
at which the nucleation rates of the two phases are the same,
as a function of $h_3-h_2$. In the case of the solid curve
the two rates equal $10^{-6}\nu$ whereas for the dotted curve
the rate equals $10^{-8}\nu$.
The impurity concentration $\rho=10^{-6}$.
\label{step}
}
\vspace*{0.3in}
\begin{center}
\epsfig{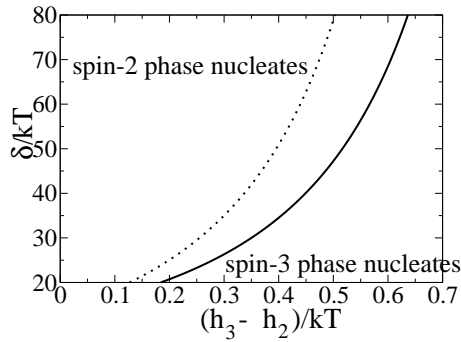}
\end{center}
\end{figure}

Let us consider systems where the difference $h_3-h_2$ is fixed,
so the spin-three phase is a fixed amount more stable.
Then $h_3-h_2$ is one parameter, $\delta$ is the other.
Having fixed both these parameters we can start with
$h_2=h_3=0$ and then increase $h_2$ and $h_3$
in parallel until the nucleation rate 
of either the spin-two, equation (\ref{r2}), or the spin-three,
equation (\ref{r3}), phase
becomes appreciable. The phase whose nucleation rate is the first
to become appreciable will then be the one that appears.
It is
a little arbitrary what nucleation rate we consider to be appreciable,
we select values of $10^{-6}$ and $10^{-8}\nu$
per site as trial values. We can then divide the
parameter space of systems in the $(h_3-h_2)$--$\delta$ plane
into a region where on increasing $h_3$ and $h_2$ at fixed $h_3-h_2$
the rate of nucleation of the spin-two phase equals $10^{-6}$
or $10^{-8}\nu$
first or a region where that of the spin-three phase
is the first to hit one of these values. We have done so and plotted
the results in figure \ref{step}; the
solid and dashed curves separate the two regions for the
different nucleation rates.
We see that as $\delta$ increases the spin-two phase appears first
even for larger and larger values of $h_3-h_2$, i.e., even when
the spin-three phase is more and more stable relative
to the spin-two phase. Impurities
that strongly favour the metastable spin-two phase,
i.e., ones with large values of $\delta$, result in the
spin-two phase preempting the spin-three phase even when the
spin-three phase is significantly more stable. This is true
even for very low impurity densities $\rho$.
They cause
this less stable phase to appear.
Note that in figure \ref{step}, $\delta\ge 20$ so our neglect
of homogeneous nucleation is always reasonable as the
rate of homogeneous nucleation as at least a factor $10^6\exp(-20)\ll 1$
smaller than the rate of heterogeneous nucleation.
Also, the trend seen in figure \ref{step} from
nucleation of the spin-two phase to nucleation of the spin-three
phase as $h_3-h_2$ is increased is just the same as that in the
results of table \ref{table1}. A given impurity
only controls nucleation, in the sense of causing the metastable
phase to appear, if the difference in stability between the
metastable phase and the equilibrium phase is not too great.

Note that for the Potts model, formation of the spin-two
phase slows down formation of the spin-three phase. This is
because the interfacial tension between the spin-two and
the spin-three phase will be very similar to that between
the spin-two and spin-three phases, both will be $\simeq J$
at low temperatures we are working at,
but the driving force for nucleation
from the spin-two phase is only $h_3-h_2$ not $h_3$.
In other systems, for example alkanes \cite{sirota99}, formation
of a metastable phase may accelerate the formation of the
equilibrium phase. See \cite{sear02} for a theoretical
description of that effect.

\section{Conclusion}

It has been known since
Ostwald's time in the nineteenth century
that when a phase has a higher free energy than two or more other
phases it is often not the equilibrium one of these other phases that
appears but a metastable one \cite{kingery}.
This was based on experimental observations.
Here we have used computer simulation
and a simple theory, and seen that which phase appears
can be controlled by the impurities that are present.
It may be that a system contains impurities that strongly favour a metastable
phase. If so, then the
metastable phase may nucleate 
on these impurities under conditions where the
equilibrium phase does not nucleate, because the barrier
to nucleation of the equilibrium phase is too high.
Thus, if we are to understand why a system obeys Ostwald's
rule, we may need to consider the effect of the impurities
present. It should be noted that the usual explanation for
Ostwald's rule, that the interfacial tension for the
metastable phase is lower than for the equilibrium phase,
is not applicable here: these two interfacial tensions will be
very similar for the low-temperature Potts model.

Here, we used Monte Carlo simulation to study heterogeneous
nucleation. Currently, experimental data is interpreted using the classical
nucleation theory of heterogeneous nucleation \cite{kingery,nucon},
but this is often unsatisfactory, see for example
\cite{cantor}. Some of the assumptions that underlie
the classical nucleation theory of heterogeneous nucleation
are known to be poor, particularly when the impurity
strongly attracts the new phase \cite{cantor}.
Thus computer simulation, which does not make these assumptions,
is useful. Experimental systems can be mapped onto the
current simple model or generalisations of it
if the supersaturations and interfacial
tensions are known. Even if the impurities in the experimental
system are uncharacterised, then simulations can be performed
with a range of impurities in order to make plausible
estimates of how strongly the impurities
in the experimental
system are interacting with the nuclei.

It is a pleasure to acknowledge
discussions with M. Dijkstra, R. Evans and D. Frenkel.



\begin{thebibliography}{99}

\bibitem{kingery} Kingery W D, Bowen H K and Uhlmann D R 1976
{\it Introduction to Ceramics} (Wiley, New York)

\bibitem{wolde99} Ten Wolde P R and Frenkel D 1999
{\it Phys. Chem. Chem. Phys.} {\bf 1} 2191

\bibitem{debenedetti} Debenedetti P G 1996
{\it Metastable Liquids}
(Princeton University Press, Princeton)

\bibitem{nucon} Proceedings of a Royal Society discussion
entitled `Nucleation Control' published as 2003
{\it Phil. Trans. A: Math. Phys. Roy. Soc. Eng. Sci.} {\bf 361},
issue 1804.
For heterogeneous nucleation see in particular
\cite{cantor,perepezko}.

\bibitem{cantor} Cantor B 2003
{\it Phil. Trans. Roy. Soc. A: Math. Phys. Eng. Sci.} {\bf 361} 409

\bibitem{perepezko} Perepezko J H and Tong W S 2003
{\it Phil. Trans. Roy. Soc. A: Math. Phys. Eng. Sci.} {\bf 361} 447

\bibitem{wu82} Wu F Y 1982 {\it Rev. Mod. Phys.} {\bf 54} 235

\bibitem{janke97} Janke W and Villanova R 1997
{\it Nucl. Phys. B} {\bf 489} 679

\bibitem{cacciuto04} Cacciuto A, Auer S and Frenkel D 2004
{\it Nature} {\bf 428} 404

\bibitem{toxvaerd02} Toxvaerd S 2002
{\it J. Chem. Phys.} {\bf 117} 10303

\bibitem{schick} Schick M 1990 in {\em Liquids at interfaces,
Les Houches XLVIII} edited by Charvolin J, Joanny J F and
Zinn-Justin J (Elsevier, Amsterdam)

\bibitem{bonn01} Bonn D and Ross D 2001
{\it Rep. Prog. Phys.} {\bf 64} 1085

\bibitem{chandler} Chandler D 1987
{\it Introduction to modern statistical mechanics}
(Oxford University Press, New York)

\bibitem{variant} When one solid nucleates on a
defect in another solid phase, the free energy
of the nucleus, and hence of course the nucleation barrier,
will depend on the orientation of the lattice planes
in the crystalline nucleus with respect both
to the phase it is nucleating in and the orientation of the defect.
Our simple Potts model with $h_2=h_3$ is a simple model
of a solid that can nucleate in only two discrete orientations
with respect to the parent phase. Then a $p=2$ impurity is
a simple model of a defect that favours the spin-two `orientation'.
Defect controlled orientation of the nucleating phase 
has been observed in experiment,
see for example
the work of Furuhara and Maki \cite{furuhara01}
who find that the orientation
of the defects controls the orientation of the face-centred-cubic
phase that nucleates in a body-centred-cubic phase of a
titanium alloy.

\bibitem{stranski33} Stranski I N and Totomanow D 1933
{\it Z. Phys. Chem.} {\bf 163} 399

\bibitem{wonczak00} Wonczak S, Strey R and Stauffer D 2000
{\it J. Chem. Phys.} {\bf 113} 1976

\bibitem{sirota99} Sirota E B and Herhold A B 1999
{\it Science} {\bf 283} 529

\bibitem{sear02} Sear R P 2002
{\it J. Phys.: Cond. Matt.} {\bf 14} 3693

\bibitem{furuhara01} Furuhara T and Maki T 2001
{\it Mat. Sci. Eng. A}  {\bf 312} 145







\end{thebibliography}
\end{document}